\documentclass[prl,twocolumn,superscriptaddress,longbibliography]{revtex4-1}
\usepackage{amssymb,amsmath,amsfonts,bm}
\usepackage{graphicx}
\usepackage{epstopdf}
\usepackage{times}
\usepackage{float}
\usepackage{lipsum}
\usepackage{color}

\newtheorem{theorem}{Theorem}

\usepackage{hyperref}
\hypersetup{  colorlinks=true, linkcolor=blue, citecolor=red, urlcolor=blue  }

\usepackage{physics}

\begin{document}

\title{Plug-and-Play Approach to Non-adiabatic Geometric Quantum Gates}


\author{Bao-Jie Liu}

\author{Xue-Ke Song}
\affiliation{Institute for Quantum Science and Engineering, and Department of Physics,
Southern University of Science and Technology, Shenzhen 518055, China}

\author{Zheng-Yuan Xue} \email{zyxue83@163.com}
\affiliation{Guangdong Provincial Key Laboratory of Quantum Engineering and Quantum Materials, GPETR Center for\\ Quantum Precision Measurement, and SPTE, South China Normal University, Guangzhou 510006, China}
\affiliation{Institute for Quantum Science and Engineering, and Department of Physics,
Southern University of Science and Technology, Shenzhen 518055, China}

\author{Xin Wang} \email{x.wang@cityu.edu.hk}
\affiliation{Department of Physics, City University of Hong Kong, Tat Chee Avenue, Kowloon, Hong Kong SAR, China, and City University of Hong Kong Shenzhen Research Institute, Shenzhen, Guangdong 518057, China}

\author{Man-Hong Yung}  \email{yung@sustc.edu.cn}
\affiliation{Institute for Quantum Science and Engineering, and Department of Physics,
Southern University of Science and Technology, Shenzhen 518055, China}
\affiliation{Shenzhen Key Laboratory of Quantum Science and Engineering, Shenzhen 518055, China}
\affiliation{Central Research Institute, Huawei Technologies, Shenzhen 518129, China}

\date{\today}

\begin{abstract}
Non-adiabatic holonomic quantum computation (NHQC) has been developed to shorten the construction times of geometric quantum gates. However, previous NHQC gates require the driving Hamiltonian to satisfy a set of rather restrictive conditions, reducing the robustness of the resulting geometric gates against control errors. Here we show that non-adiabatic geometric gates can be constructed in an extensible way, called NHQC+, for maintaining both flexibility and robustness against certain types of noises. Consequently, this approach makes it possible to incorporate most of the existing optimal control methods, such as dynamical decoupling, composite pulses, and shortcut to adiabaticity, into the construction of single-looped geometric gates. Furthermore, this extensible approach of geometric quantum computation can be applied to various physical platforms such as superconducting qubits and nitrogen-vacancy centers. Specifically, we performed numerical simulation to show how the noise robustness in the recent experimental implementations [Phys. Rev. Lett. 119, 140503 (2017)] and [Nat. Photonics 11, 309 (2017)] can be significantly improved by our NHQC+ approach. These results cover a large class of new techniques combing the noise robustness of both geometric phase and optimal control theory.
\end{abstract}


\maketitle

\emph{\bf Introduction}.\textbf{--} Methods of constructing precise and noise-resistant quantum gates are of  fundamental importance to quantum information processing. Geometric quantum computation (GQC) 
utilizes a peculiar property of quantum theory: the Abelian or non-Abelian {\it geometric phases} \cite{Zanardi1999,gqc,b1} of  quantum states, acquired after a cyclic evolution. The geometric phase depends only on the global properties of the evolution trajectories. Consequently, quantum gates based on the geometric phases are immune to local disturbances during the evolution~\cite{Zhu2005,Berger2013,Chiara,Leek,Filipp}. More precisely, a geometric phase can either be a real number (i.e., Abelian), known as the "Berry phase"~\cite{b2},  or a matrix (non-Abelian holonomy)~\cite{b3} that is the key ingredient in constructing quantum operations for holonomic quantum computation (HQC).

Early applications of GQC involves {\it adiabatic} evolutions  to avoid transitions between different sets of eigenstates \cite{Jones,Duan2001a,lian2005}. However, the adiabatic condition necessarily implies lengthy gate operation time; the effectiveness of adiabatic GQC therefore becomes severely limited by the environment-induced decoherence. Later, it was found that geometric quantum gates can be realized {\it non-adiabatically}, if we construct a driving Hamiltonian with {time-independent} eigenstates~\cite{b4,b5,b6,Sjoqvist2012,Xu2012,xue2017,Xu2018}, and build geometric gates confined to the Hilbert subspace~\cite{Sjoqvist2012,Xu2012}, a method known as NHQC (non-adiabatic holonomic quantum computation). However, this condition imposes stringent requirements on the driving Hamiltonian; the systematic errors would introduce additional fluctuating phase shifts, smearing the geometric phases~\cite{Thomas,Johansson,Zheng,jun2017}. Furthermore, the restriction imposed in the previous NHQC schemes excludes the flexibility for incorporating most of the optimization techniques, limiting its applicability. These problems motivate us to search for a new approach to GQC that is (i) non-adiabatic, (ii) robust against  the control errors, (iii) compatible with other optimization techniques for maximizing the gate fidelity against different types of noises.

\begin{figure}[tbp]
\centering\includegraphics[width=8.5cm]{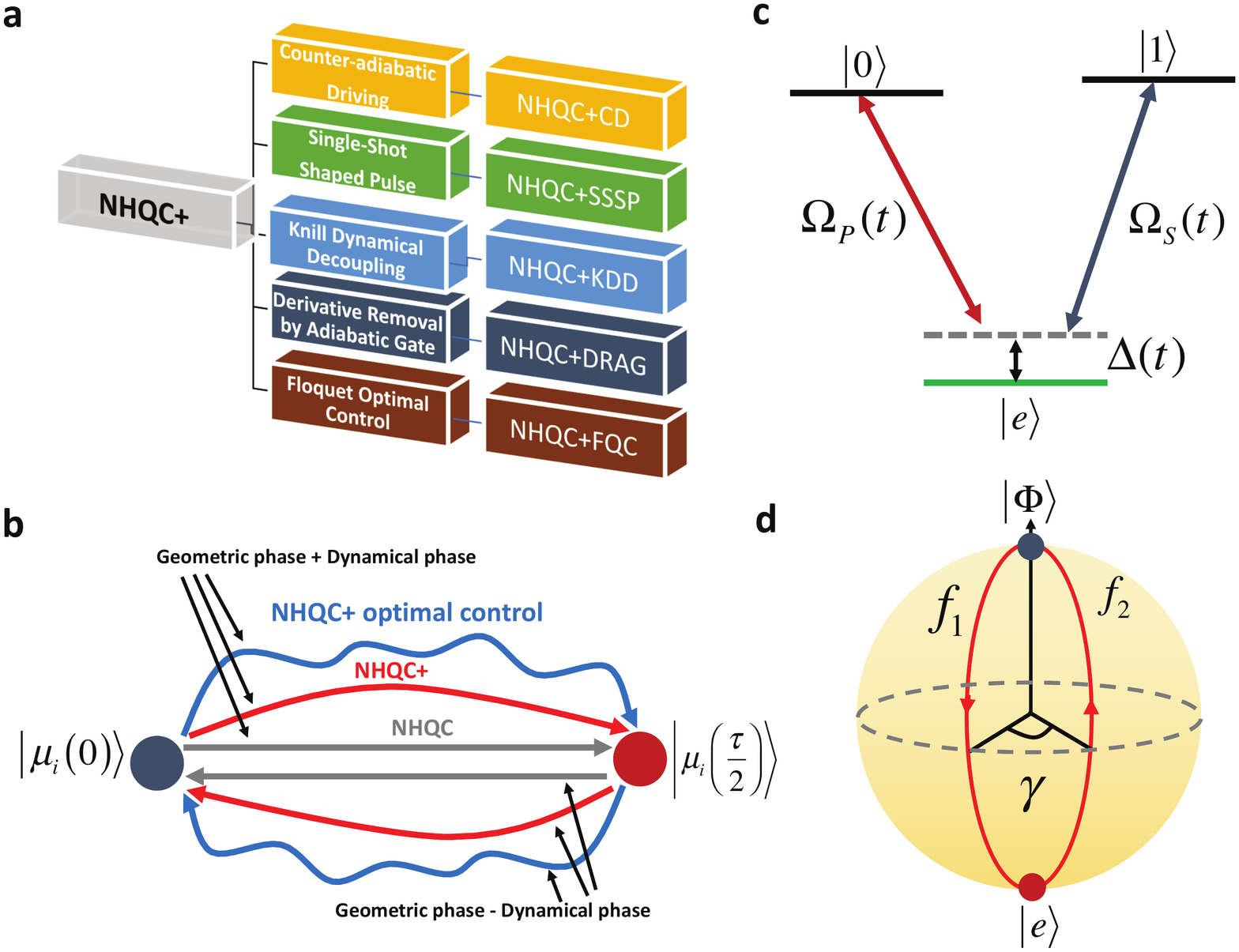}\caption{\label{setup} The illustration of our proposed implementation. (a) Schematic of combining various optimal control pulses with geometric quantum computation in our scheme. (b) Schematic of different  path for GQC: NHQC (gray), NHQC+ (red) is compatible with optimization methods with choosing
different paths as NHQC+ optimal control (blue). (c) The level structure and coupling configuration for single-qubit operations with a solid-state NV center, where driven pulses with amplitudes $\Omega_{P}$ and $\Omega_{S}$ couple $|0\rangle$ and $|1\rangle$ to $|e\rangle$ with detuning $\Delta$. (d) Conceptual explanation for geometric quantum operation.
}
\end{figure}

Here, we demonstrate that non-adiabatic GQC is also possible under much general conditions, relative to traditional NHQC \cite{Sjoqvist2012,Xu2012}. Our approach leads to a new form of single-looped GQC that is compatible with most of the existing pulse-shape optimization methods, including Derivative Removal by Adiabatic Gate (DRAG) \cite{DRAG}, counteradiabatic driving (CD)~\cite{Berry,chenxi,Zhang2015Srep,Song2016NJP,Liu2017PRA,Yan2019}, dynamical decoupling (DD) \cite{DD1,DD2,DD3},  dynamically-corrected gates (DCG) \cite{Kaveh,BB1,SUPCODE}, Floquet optimal control~\cite{Bartels2011pra,bauer}, and machine-learning-based optimization techniques~\cite{zhang2018,yang2018}, etc, as shown in Fig \ref{setup}a. Given the extensibility of this approach and the fact that the traditional NHQC method can be regarded as a special case, we refer to this method as NHQC+~\cite{footnote}. For example, when combined with CD, we label it as NHQC+CD.

In the following, we shall employ a three-level quantum systems to illustrate the working mechanism of NHQC+. In particular, we are interested in comparing our method with the NHQC gates implemented in recent experiments with NV centers~\cite{s0,s2}. Numerical simulations indicate that our optimized NHQC+ method can achieve a significant improvement over the NHQC gates in Refs.~\cite{s0,s2}, using the experimental parameters.

\emph{\bf General framework}.{---} Let us start with a general time-dependent Hamiltonian~$H(t)$. For any complete set of basis vectors, $\left\{ {\left| {{\psi _m}\left( 0 \right)} \right\rangle } \right\}$ at $t=0$, $U\left( t,0 \right) = {\mathcal T}{e^{ - i\int_0^t {H\left( {t'} \right)} dt'}} = \sum\nolimits_m {\left| {{\psi _m}\left( t \right)} \right\rangle \left\langle {{\psi _m}\left( 0 \right)} \right|}$,
where the time-dependent state,  $\left| {{\psi _m}\left( t \right)} \right\rangle  = {\mathcal T}{e^{ - i\int_0^t {H\left( {t'} \right)} dt'}}\left| {{\psi _m}\left( 0 \right)} \right\rangle $, follows the Schr\"{o}dinger equation. Now, at each moment of time, we can always choose a different set of time-dependent basis, $\left\{ {\left| {{\mu _m}\left( t \right)} \right\rangle } \right\}$, which satisfies the boundary conditions at time $t=0$ and $t=\tau$:
\begin{equation}\label{boundcon}
\left| {{\mu _m}\left( \tau  \right)} \right\rangle  = \left| {{\mu _m}\left( 0 \right)} \right\rangle  = \left| {{\psi _m}\left( 0 \right)} \right\rangle \ ,
\end{equation}
but in general their time dependence do not  follow Schr\"{o}dinger's equation. In this way, we can always write, $\left| {{\psi _m}\left( t \right)} \right\rangle  = \sum\nolimits_k {{v_{km}}\left( t \right)} \left| {{\mu _k}\left( t \right)} \right\rangle$,
which means that the time-evolution operator becomes $U\left( t,0 \right) = \sum\nolimits_{m,k} {{v_{km}}\left( t \right)\left| {{\mu _k}\left( t \right)} \right\rangle \left\langle {{\psi _m}\left( 0 \right)} \right|} $. Applying the boundary conditions, we obtain the following unitary transformation matrix at the final time $t=\tau$, $U\left( \tau,0  \right) = \sum\nolimits_{m,k} {{v_{km}}\left( \tau  \right)\left| {{\psi _k}\left( 0 \right)} \right\rangle \left\langle {{\psi _m}\left( 0 \right)} \right|}$.
The matrix element~${{v_{mk}}\left( \tau  \right)}$ satisfies the following equation,
\begin{equation}\label{dvAHv}
\frac{d}{{dt}}{v_{km}}\left( t \right) = i\sum\limits_{l = 1} {\left( {{A_{kl}}\left( t \right) - {H_{kl}}\left( t \right)} \right)} {v_{lm}}\left( t \right) \ ,
\end{equation}
where ${H_{kl}}\left( t \right) \equiv \left\langle {{\mu _k}\left( t \right)} \right|H\left( t \right)\left| {{\mu _l}\left( t \right)} \right\rangle$ and ${A_{kl}}\left( t \right) \equiv i\left\langle {{\mu _k}\left( t \right)} \right|\frac{d}{{dt}}\left| {{\mu _l}\left( t \right)} \right\rangle$, which can be combined to form an effective Hamiltonian: ${H_{\rm eff}}\left( t \right) \equiv V(t)^\dagger H\left( t \right)V(t) - i{V^\dag }\left( t \right)\frac{d}{{dt}}V\left( t \right)$, where $V(t) \equiv \sum_{k}|\mu_k(t)\rangle\langle \mu_k(0)|$. In other words, written in the initial basis $\left\{ {\left| {{\mu _k}\left( 0 \right)} \right\rangle } \right\}$, the matrix elements are given by ${A_{kl}}\left( t \right)-{H_{kl}}\left( t \right)$. With these tools, various forms of GQC can emerge as different settings (or approximations) of these equations (see SM~\cite{Supp}).

\emph{\bf Conditions of NHQC+}.{---} Our strategy is to find an auxiliary  basis $\left\{ {\left| {{\mu _k}\left( t \right)} \right\rangle } \right\}$ such that for all $k \ne m$, the effective Hamiltonian $H_{\rm eff}$ is always diagonal in the {\it initial basis}, i.e.,
\begin{equation}\label{phiHeffphimk}
\left\langle {{\mu _m}\left( 0 \right)} \right|{H_{\rm eff}}\left( t \right)\left| {{\mu _k}\left( 0 \right)} \right\rangle  = 0 \ .
\end{equation}
Consequently, Eq.~(\ref{dvAHv}) implies that ${v_{mk}}(t) = {\delta _{mk}} \ {v_{kk}}\left( t \right)$ is also diagonal and hence the unitary operator,
\begin{equation}\label{Uvmmk}
U\left( t,0  \right) = \sum\nolimits_{k} {{v_{kk}}\left( t  \right)\left| {{\mu _k}\left( t \right)} \right\rangle \left\langle {{\mu _k}\left( 0 \right)} \right|}
\end{equation}
are all diagonal, where ${v_{kk}}\left( t \right) = {e^{ - i\int_0^\tau  {\left\langle {{\mu _k}\left( t \right)} \right|H\left( t \right)} \left| {{\mu _k}\left( t \right)} \right\rangle dt - \int_0^\tau  {\left\langle {{\mu _k}\left( t \right)} \right|} \left. {{{\dot \mu }_k}\left( t \right)} \right\rangle dt}}$. Note particularly that if the following condition,
\begin{equation}
\label{npipgpz}
\int_{0}^{\tau}\langle\mu_{k}(t)|H(t)|\mu_{k}(t)\rangle \ dt=0.
\end{equation}
is further satisfied for {\it each} $k$, then the resulting unitary evolution becomes {\it purely geometric}, i.e.,
\begin{equation}
U\left( \tau,0  \right) = \sum\limits_k {{e^{ - \int_0^\tau  {\langle {\mu _k}(t)|{{\dot \mu }_k}(t)\rangle } dt}}\left| {{\mu _k}\left( 0 \right)} \right\rangle \langle {\mu _k}\left( 0 \right)|} \ ,
\end{equation}
which is the main goal that can be achieved through the following theorem:

\begin{theorem}[NHQC+ equation] The condition in Eq.~(\ref{phiHeffphimk}) is satisfied  only if the Hamiltonian $H(t)$ and the projector ${\Pi _k}(t) \equiv \left| {{\mu _k}\left( t \right)} \right\rangle \left\langle {{\mu _k}\left( t \right)} \right|$ of the auxiliary  basis $\left\{ {\left| {{\mu _k}\left( t \right)} \right\rangle } \right\}$ follows the von Neumann equation, i.e.,
\begin{equation}\label{invarint}
\frac{{d}}{{dt}} \Pi_{k}(t)={-i}\left[ {H(t),\Pi_{k}(t)} \right]\ ,
\end{equation}
\end{theorem}
The proof of this theorem is given in SM~\cite{Supp}.

Note that the key difference between the  previous NHQC schemes and the NHQC+ approach  introduced here is that the Hamiltonians are subject to different constraints. In the NHQC case, the Hamiltonian is required to satisfy a set of constraints: $\left\langle {{\psi _{m}}\left( t \right)} \right|H\left( t \right)\left| {{\psi _{k}}\left( t \right)} \right\rangle  = 0$, which is required (i) at {\it each moment of time} and (ii) for {\it all} possible $k,m$. However, for NHQC+, the Hamiltonian needs to vanish only in the integral sense (see Eq.~(\ref{npipgpz})). More importantly, the NHQC+ removes the constraints for $k\neq m$, which makes it possible for our method being compatible with most of the optimization schemes (see  Fig.~\ref{setup}a).

\emph{\bf Application of NHQC+ gates}.\textbf{---} To continue, we shall make our discussion explicit by demonstrating its application in realistic systems, namely NV center.  Specifically, we shall focus on the three-level $\Lambda$ configuration in Ref.~\cite{Zu2014} with a one-photon detuning $\Delta(t)$, as shown in Fig.~\ref{setup} c.

The spin states are labelled as $|E_{\nu}\rangle\otimes|m_s\rangle$, where $|E_{\nu}\rangle$ and $|m_s\rangle$ denote respectively the orbital and spin states. For our purpose, we choose $|0\rangle \equiv |E_{0}\rangle\otimes|1\rangle$, $|1\rangle \equiv |E_{0}\rangle\otimes|-1\rangle$  and $|2\rangle \equiv |E_{0}\rangle\otimes|0\rangle$. In this subspace, the corresponding Hamiltonian, in the interaction picture, is given by~\cite{Vitanov},
\begin{equation} \label{H0}
H(t)=\Delta(t)|e\rangle\langle e|+
\frac{1}{2}\left[\left(\Omega_{P}(t)|0\rangle+\Omega_{S}(t)|1\rangle\right)\langle e|+H.c.\right],
\end{equation}
where $\Omega_{P}(t)$ and $\Omega_{S}(t)$ denote, respectively, the pumping and Stokes pulses driving the $|0\rangle\leftrightarrow|e\rangle$ and $|1\rangle\leftrightarrow|e\rangle$ transitions.

Here, we choose the pluses to have the following form, $\Omega_{P}(t)=\Omega(t)\sin(\theta/2)e^{i\phi_{1}(t)}$ and $\Omega_{S}(t)=\Omega(t)\cos(\theta/2)e^{i[\phi_{1}(t)+\phi]}$, but we maintain the ratio $\Omega_{P}(t)/\Omega_{S}(t)$  of the two pulses to be time-independent, i.e., ${\Omega _P}\left( t \right)/{\Omega _S}\left( t \right) \equiv \tan \left( {\theta /2} \right){e^{ - i\phi }}$. Consequently, the Hamiltonian in Eq. (\ref{H0}) can be simplified as,
\begin{equation}\label{CtHiEsim}
{H}(t) = \Delta (t)|e\rangle \langle e| + \frac{{\Omega (t)}}{2}\left[ {e^{i\phi_{1}(t)}\left| \Phi  \right\rangle \langle e| + H.c.} \right] \ ,
\end{equation}
where we defined a time-independent bright state as, $|\Phi\rangle \equiv\sin(\theta/2)|0\rangle+\cos(\theta/2)e^{i\phi}|1\rangle$.

Recall that for realizing NHQC+ gates, we need to choose a set of auxiliary states satisfying the boundary conditions in Eq.~(\ref{boundcon}). Here our choice is (i) a dark state $|\mu_{0}\rangle =\cos(\theta/2)|0\rangle-\sin(\theta/2)e^{i\phi}|1\rangle$ which is decoupled from subspace of $|\Phi\rangle$ and $|e\rangle$, and (ii) an orthogonal state in the following form:
\begin{eqnarray}
\label{esotdi3}
|{\mu _ + }(t)\rangle  = \sin \tfrac{{\chi (t)}}{2}|\Phi \rangle  + \cos \tfrac{{\chi (t)}}{2}{e^{-i\alpha (t)}}|e\rangle  \ .
\end{eqnarray}
The variables $\chi(t)$ and $\alpha(t)$ are determined by requiring the corresponding projector, $\Pi_{+}(t)=|\mu_{+}(t)\rangle\langle\mu_{+}(t)|$, to satisfy the von Neumann equation in Eq.~(\ref{invarint}). Explicitly, we found that they are governed by the following coupled differential equations:
\begin{align}\label{I33}
\Omega(t) & = \frac{\dot{\chi}(t)}{\sin[\phi_{1}(t)-\alpha(t)]} \nonumber \ , \\  \Delta(t) & =-\dot{\alpha}(t) -\dot{\chi}(t)\cot\chi(t)\cot[\phi_{1}(t)-\alpha(t)] \ .
\end{align}
We have a lot of choices to pick the variables $\chi(t)$ and $\alpha(t)$ as long as they meet the coupled equations Eq. (\ref{I33}) and the cyclic evolution conditions Eq. (\ref{boundcon}). One possible set of solution is found to be $\chi (t) = \pi [{\text{erf}}(2t/T) + 1]$ and $\alpha (t) = {\phi _1}(t) - \arctan [\dot \chi (t)/({\Omega _0}\sin \chi (t))]$, where the parameter $T$ controls the effective duration of the protocol. Since experimentally the driving pulses have finite duration, we choose $\tau=4T$ as a cutoff. In this way, the time dependence of $\Omega(t)$ and $\Delta(t)$ can be determined numerically.

\emph{\bf Construction of NHQC+ gates}.\textbf{---} We are now ready to demonstrate how to build up universal non-Abelian geometric single-qubit gates, i.e., holonomic quantum gates. Let us start with the following set of basis states, $\left\{ {\left| {{\mu _0}} \right\rangle },{\left| {{\mu _+}\left( 0 \right)} \right\rangle } \right\}$. Note that this basis is only spanned by the computational basis states $\left\{|0\rangle,|1\rangle\right\}$, as $\chi(0)=\pi$. Next, we consider that the physical procedure is divided into two time intervals $(0,\tau/2)$ and $(\tau/2,\tau)$. During the first interval ($0\le t\le\tau/2$), the phase angle $\phi_{1}(t)$ of the microwave is set to be a constant~$\gamma_{1}$, i.e., $\phi_{1}(t)=\gamma_{1}$. From Eq.~(\ref{Uvmmk}), the corresponding evolution operator is given by, ${U_1}(\tau/2,0) = |{\mu _0}\rangle \langle {\mu _0}| + {e^{i{f _1}}}|{\mu _ + }(\tau /2)\rangle \langle {\mu _ + }(0)|$, where ${f _1} \equiv - f \left( {\tau /2,0} \right)$.
During the second interval ($\tau/2\le t\le\tau$), the phase $\phi_{1}(t)$ is chosen to be another constant, $\phi_{1} (t)=\gamma_{2}$. In the gate implementation, the dynamical phases will also be acquired for both intervals. To remove the total dynamical phase, we intend to set the dynamical phase of the latter interval to be the opposite to the formal one, as shown in Fig. \ref{setup}b, and thus the total dynamical phase will be sum to zero. This can be achieved by the spin-echo technique~\cite{Jones}, as we will explicitly investigate later, and the pulse-shaping method \cite{Supp}, which will lead to gates solely dependent on geometric phases of the cyclic states.
As a result, the corresponding evolution operator takes the form, $U_2(\tau,\tau/2)=|\mu_{0}\rangle\langle \mu_{0}|+e^{if_{2}}|\mu_+(\tau)\rangle\langle \mu_+(\tau /2)|$, where we defined another phase factor ${f _2} \equiv - f \left( \tau,\tau/2  \right)$.

Finally, applying the boundary condition Eq.~(\ref{boundcon}), we obtain the following unitary transformation matrix in the basis states $\left\{ {\left| {{\mu _0}} \right\rangle },{\left| {{\mu _+}\left( 0 \right)} \right\rangle } \right\}$ at the final time $t=\tau$,
\begin{equation}\label{U1}
U(\tau,0 ) = {U_2}{U_1} = \left| {{\mu _0}} \right\rangle \left\langle {{\mu _0}} \right| + {e^{i\gamma }}\left| {{\mu _ + }\left( 0 \right)} \right\rangle \left\langle {{\mu _ + }\left( 0 \right)} \right| \ ,
\end{equation}
where the phase factor, $\gamma={{f_{1}}}+{{f_{2}}}$, is purely a geometric phase, as the condition in Eq.~(\ref{npipgpz}) has been satisfied for eliminating the dynamical phase.

\begin{figure}[tbp]
\centering
\includegraphics[width=8.5cm,height=4.0cm]{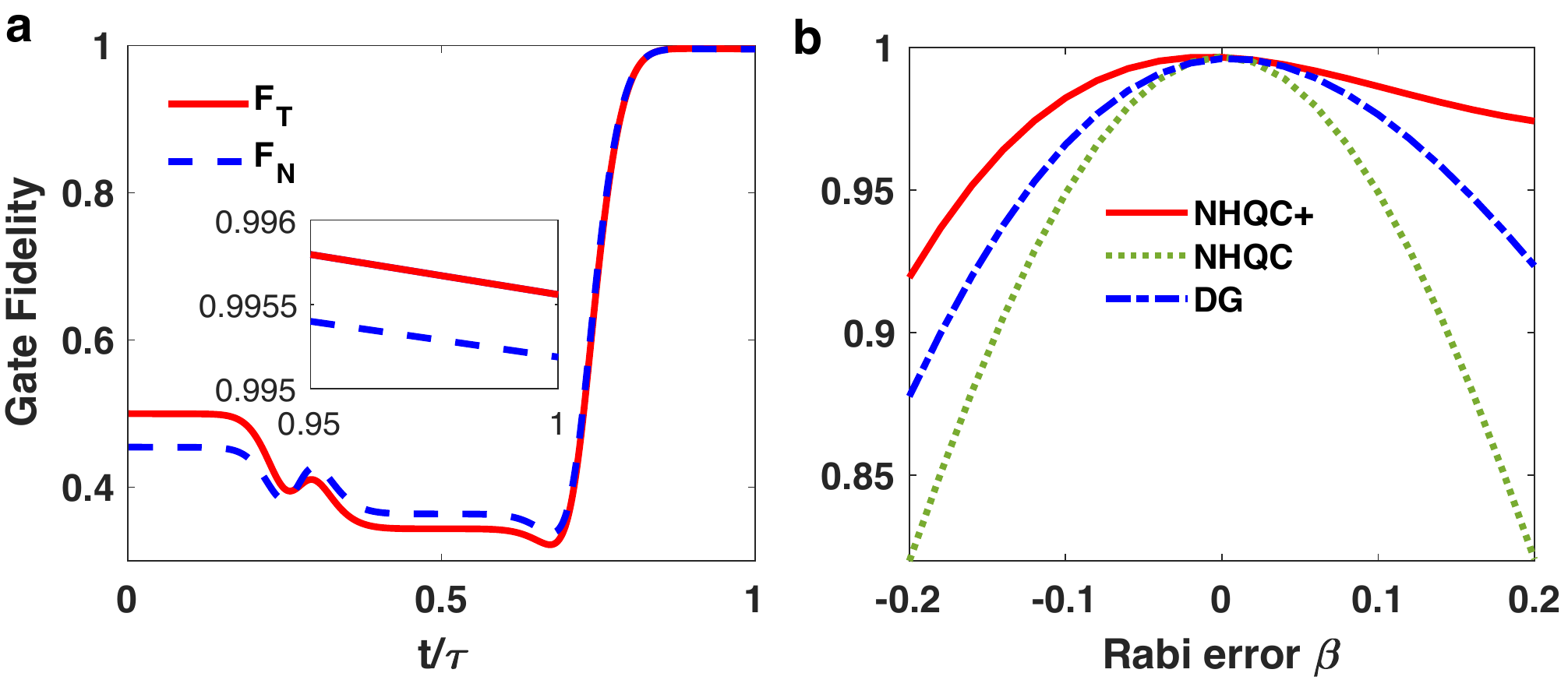}
\caption{(a) The dynamics of T gate and NOT gate fidelities, as a function of $t/\tau$. (b) The NOT gate robustness of NHQC+, NHQC and DG  as a function of systemic error of Rabi frequency, i.e., the relative pulse deviation $\beta$.}\label{single}
\end{figure}

In fact, one can show that the geometric phase $\gamma$ is exactly equal to half of the solid angle $\Omega_{\rm angle}$ shown in Fig \ref{setup} d, i.e., $\gamma = \Omega_{\rm angle}/2$. Considering two stages, a closed path in the parameter space is formed. The solid angle enclosed by the closed path is evidently $2(\gamma_{2}-\gamma_{1})$ and thus, the geometric phase acquired is simply $\gamma=\gamma_{2}-\gamma_{1}$. Therefore, the generic resilience of NHQC+ is the same as that of the conventional NHQC.

Note that the non-Abelian geometric gate $U(\tau,0)$ can be spanned by the logical $\{|0\rangle,|1\rangle\}$ basis, i.e.,
\begin{eqnarray}\label{U2}
U_{\gamma,\theta,\phi}
=e^{i\frac{\gamma}{2}}e^{-i\frac{\gamma}{2}\bm{n}\cdot\bm{\sigma}},
\end{eqnarray}
where $\bm{n}=(\sin\theta\cos\phi,\sin\theta\sin\phi,\cos\theta)$, $\bm{\sigma}$ are the Pauli matrices.  Eq. (\ref{U2}) describes a rotational operation around the $\bm{n}$ axis by a $\gamma$ angle, up to a global phase factor $e^{-i\frac{\gamma}{2}}$. As both $\bm{n}$ and $\gamma$ can take any value,  Eq. (\ref{U2}) denotes a set of universal single-qubit gates in the qubit subspace.

\begin{figure*}[tbp]
\centering
\includegraphics[width=15.5cm,height=4.0cm]{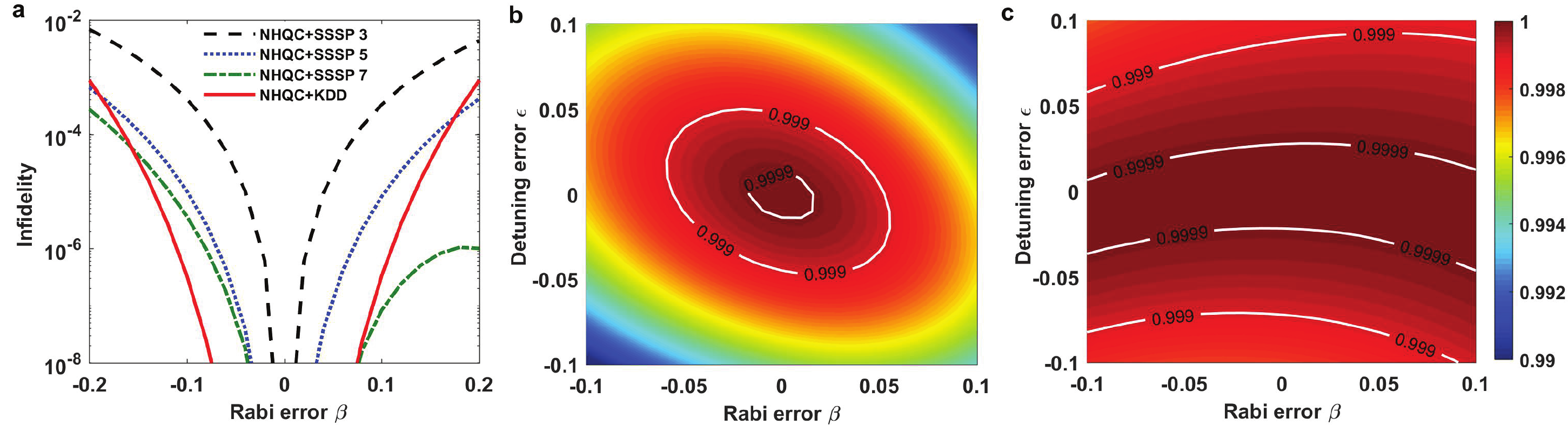}
\caption{ The performance of the gates against pulse errors without considering relaxation. (a) The T gate robustness of different orders of SSSP and KDD pulse as a function of Rabi error $\beta$. The NOT gate fidelities of (b) NHQC and (c) NHQC+SSSP 3 as a functions of $\beta$ and $\varepsilon$. Paremeters for: NHQC+ KDD $\tau_{\pi}=0.01\tau$; NHQC+SSSP3: $\{a_{1}=-1, a_{2}=a_{3}=0$\}; SSSP5: $\{a_{1}=-2.4864, a_{2}=-0.74, a_{3}=0$\}; SSSP7: $\{a_{1}=-3.46, a_{2}=-1.365, a_{3}=-0.5$\}.}\label{Dephasing}
\end{figure*}

The performance of the geometric gate in Eq.~(\ref{U2}) can be simulated by using the Lindblad master equation~\cite{Supp,Lindblad}. In our simulation, we have used the following set of experimental parameters~\cite{RPP}: the gate time is set as, $\tau=120$ ns, and  the decay and dephasing rates of the NV center are taken as $\Gamma_1^j=\Gamma_2^j \approx 2\pi \times 10$ KHz. We have investigated the gate fidelity of the T and NOT gates for initial states of the form $|\psi\rangle=\cos\Theta|0\rangle+\sin\Theta|1\rangle$, where a total of 1001 different values of $\Theta$ were  uniformly sampled in the range of $[0, \pi/2]$, as shown in Fig. \ref{single}a. The gate fidelities of the T ($U_{\pi/4, 0, 0}$) and NOT ($U_{\pi, \pi/2, 0}$) gates can reach, respectively, $99.56\%$ and $99.53\%$.

The method is not only robust against slow Rabi errors and random quasi-static noises, but also compatible with various pulse optimization techniques to further enhance the robustness against different types of noises, for example, slow Rabi control and dephasing errors, and decoherence. Explicitly, we investigate the robustness of NHQC+ against pulse errors caused by a usual slow quasistatic noise~\cite{BB1,bauer}. Here we allow the driving pulse to vary in strength, i.e., ${\Omega _{\max }} \to (1 + \beta ){\Omega _{\max }}$, where $\beta\in[-0.2,0.2]$ represents the Rabi error. In other words, the Hamiltonian becomes $H(t) \to H(t)+ H_{\beta}(t)$, where $H_{\beta}(t)=\beta\Omega(t)(|\Phi\rangle\langle e|+H.c.)$ is the perturbing Hamiltonian. Comparing our NHQC+ (before optimization) with the traditional NHQC methods and dynamical gate (DG) without canceling the dynamical phase~\cite{DG}, we simulated the performance of the same non-abelian NOT gate with the same pulse error (see in SM~\cite{Supp} for these pulse shapes). As shown in Fig. \ref{single}b, the NHQC+ gate is not only always more robust than the DG gate but also NHQC gate. In addition , we also show the robustness of NHQC+ against the random quasi-static noises for different gates in the SM~\cite{Supp} since the Rabi error $\beta$ may also randomly vary between different runs of an experiment.

\emph{\bf Optimization 1: NHQC+SSSP}.\textbf{---} Here, we demonstrate that our method can be compatible with other pulse optimization techniques to further enhance the robustness against the slow Rabi error, without considering the relaxation.   The unperturbed evolution operator can be written (see Eq.~(\ref{Uvmmk})) as $U(v,t)=|\mu^{0}_{0}(v)\rangle\langle \mu^{0}_{0}(t)|+e^{if(t)}|\mu^{0}_{+}(v)\rangle\langle \mu^{0}_{+}(t)|+e^{-if(t)}|\mu^{0}_{-}(v)\rangle\langle\mu^{0}_{-}(t)|$, where $|{\mu _ - }(t)\rangle  = \cos \tfrac{{\chi (t)}}{2}|\Phi \rangle  - \sin \tfrac{{\chi (t)}}{2}{e^{-i\alpha (t)}}|e\rangle $ is an orthogonal state of $|\mu^{0}_{+}(t)\rangle$ defined in Eq.~(\ref{esotdi3}). Suppose the system is initialized in the state $|\mu_{+}^{0}(0)\rangle$ and $\beta$ is small, to the second order, we write $|\mu_{+}(\tau)\rangle = |\mu^{0}_{+}(\tau)\rangle + |\mu^{1}_{+}(\tau)\rangle + |\mu^{2}_{+}(\tau)\rangle + \emph{O}(\beta^{3}) $, where $|\mu^{0}_{+}(t)\rangle$ is the unperturbed term, $|\mu^{1}_{+}(\tau)\rangle = -i\int_{0}^{\tau}dt \ U_{0}(\tau,t)H_{\beta}(t)|\mu_{+}^{0}(t)\rangle$, and $|\mu^{2}_{+}(\tau)\rangle = - \int_{0}^{\tau}dt\int_{0}^{t}dt'U_{0}(\tau,t)H_{\beta}(t)  U_{0}(t,t')H_{\beta}(t')|\mu_{+}^{0}(t')\rangle$.

Consequently, the  fidelity or population is given by $P \equiv {\left| {\left\langle {\mu _ + ^0(\tau )} \right.|{\mu _ + }(\tau )\rangle } \right|^2} =1-\beta^{2}{\left|\int_{0}^{\tau}\langle\mu^{0}_{+}(t)|H_{\beta}(t)|\mu^{0}_{-}(t)\rangle dt\right|^2}+\emph{O}(\beta^{3}).
$ The sensitivity to the systematic error~\cite{OSTA1,Song2017} can be quantified as $Q_{s} \equiv -\frac{1}{2}\frac{\partial^{2}P(\beta)}{\partial\beta^2}|_{\beta=0}=-\frac{\partial P(\beta)}{\partial(\beta^2)}|_{\beta=0}$. Explicitly, we found that
\begin{equation}
{Q_s} = {\left| {\int_0^\tau \dot{\chi}(t) \sin^{2}{\chi}(t) e^{- if (t)}dt} \right|^2} \ .
\end{equation}
To minimize $Q_{s}$, we can choose the SSSP (single-shot shaped pulse) through $f(t)$ as a function of $\chi(t)$ with a Fourier-series type of Ansatz (see~\cite{Supp,OSTA2}), i.e., $f(t)=2\chi(t)+\sum_{n}a_{n}\sin[2n \chi(t)]$, where $\chi=\pi[\text{erf}(2t/T)+1]$. Here, the set $\{a_{n}\}$ of coefficients are chosen such that $Q_s$ is minimized, and other parameters are  determined in SM~\cite{Supp}. The performance of various choices of $\{a_{n}\}$ is shown in~Fig.~\ref{Dephasing} a, indicting that the sensitivity decreases when more terms of $a_n$ are included.

On the other hand, we also compare the T gate robustness of an optimized NHQC+SSSP with the NHQC scheme applied in the recent experiments with NV centers~\cite{s0,s2}, as shown in Fig.~\ref{Dephasing} b and c for large deviations of the original Rabi frequency $\beta$ (up to $10\%$) and a range of the static detuning $\delta$ of the same order as the peak value of the Rabi frequencies.

\emph{\bf Optimization 2: NHQC+KDD}.\textbf{---} As an another example, one can also combine NHQC+ with Knill's dynamical-decoupling (KDD) technique~\cite{DD2}, to simultaneously compensate the Rabi and dephasing induced errors. Recall that DD sequences consist of repetitive trains of $\pi$ pulses. The delay between the pulses and their phases are important parameters for the performance of the DD pulse sequences, where the robustness can be enhanced by the so-called KDD ``self-correcting" pulses~\cite{DD2} are given by, $
KDD=F_{\tau/2}-(\pi)_{\pi/6}-F_{\tau}-(\pi)_{0}-F_{\tau}-(\pi)_{\pi/2}-F_{\tau}-(\pi)_{0}-F_{\tau}-(\pi)_{\pi/6}-F_{\tau/2},$
where $F_{\tau}$ represents free evolution for a period of time $\tau$ and $(\pi)_{\eta}$ represents $\pi$ rotation along the axis $\eta$ away from the $z$-axis.

To construct a NHQC+KDD gate, we require each $\pi$ pulse to be constructed with geometrical protection~(see Eq.~\ref{U1}). Explicitly, we choose the following control sequence for our Hamiltonian~Eq.~(\ref{CtHiEsim}), $ \chi \left( t \right) =\frac { \pi  }{ 2 } \sin\left[ \left( \frac { \pi \left( t-\tau /2 \right)  }{ { \tau  }_{ \pi  } }  \right)  \right], \alpha \left( t \right) ={ \phi  }_{ 1 }\left( t \right) +\frac { \pi  }{ 2 }, \phi_{1}(t)=\eta,$ where $\tau_{\pi} \ll \tau$ represents the total time of each $\pi$ pulse. As a demonstration, we construct a holonomic $U_{\pi, \theta, \phi}$ gate with a total of 20 self-correcting pulses by combining 5 KDD blocks to form: $[KDD_{\eta}-KDD_{\eta+\pi/2}-KDD_{\eta}-KDD_{\eta+\pi/2}]$
~\cite{DD2}.
For numerical simulation, we plot the NOT gate to against Rabi errors, as shown in Fig.~\ref{Dephasing}a, which can outperform NHQC+SSSP pulses for small Rabi errors.


\emph{\bf Conclusion}.\textbf{---} We have presented an extensible framework of non-adiabatic geometric quantum computation, NHQC+, which is compatible with many techniques in optimal control theory, such as SSSP, KDD, and more (see SM~\cite{Supp}).  Our approach relaxes the constraint imposed for the driving Hamiltonian in the previous approach of NHQC. We also presented an explicit way to implement our approach for three level systems, and numerically simulated the performance of pulse optimization for nitrogen-vacancy center. In addition, we discuss how NHQC+ gate presented here can be applied to two-qubit gates in SM~\cite{Supp}. An open question for further developement is to understand the advantages and disadvantages of NHQC+, compared with alternative control schemes.

\bigskip

\acknowledgments
The authors thank J. Jing for helpful discussions. This work is supported by the National Natural Science Foundation of China (Grant No. 11604277, No. 11405093, and No. 11874156), the Key R\&D Program of Guangdong province (Grant No. 2018B030326001), the Research Grants Council of Hong Kong  (Grants No. CityU 21300116 and No. CityU 11303617), the Guangdong Innovative and Entrepreneurial Research Team Program (Grant No. 2016ZT06D348), Natural Science Foundation of Guangdong Province (Grant No. 2017B030308003), and the Science, Technology and Innovation Commission of Shenzhen Municipality (Grants No. ZDSYS20170303165926217 and No. JCYJ20170412152620376), the National Key R\&D Program of China (Grant No. 2016YFA0301803).



\begin{thebibliography}{99}
\bibitem{footnote} Note that NHQC+ is the quantum gate based on non-Abelian geometric phase, which could be exploited  for the purposes of noise-resilient quantum computation.
\bibitem{DG} For the DG gate, the condition in Eq.~(\ref{npipgpz}) wasn't satisfied because we did not apply a spin-echo puslehalfway ($t=\tau/2$) of the control sequence to eliminate the dynamical phase accumulated during the evolution; other than this, the other parameters DG and NHQC+ are the same
(see in SM~\cite{Supp}).
\end{thebibliography}
\end{document}